\documentclass[amsmath,amssymb,twocolumn]{revtex4}
\usepackage{dcolumn}
\usepackage{bm}
\usepackage{amsmath,mathrsfs}
\usepackage{graphicx,epsfig}
\usepackage{epsf}
\usepackage{color}

\begin{document}


\title{{\bf Evolving wormhole in the brane-world scenario}}

\author{ F. Parsaei $^{1}$}\email{email:fparsaei@gmail.com}
\author{N. Riazi$^{2}$}\email{n_riazi@sbu.ac.ir}
\affiliation{$1$. Physics Department , Sirjan University of Technology, Sirjan 78137, Iran.,
\\ $2$ .Physics Department, Shahid Beheshti University, Tehran
19839, Iran}

\date{\today}


\begin{abstract}
\par  In this  paper, evolving wormholes in the context of brane-world scenario are investigated. We have studied the possible dynamic solutions  with different forms of  Ricci scalar. The possibility of existence of dynamic traversable wormholes, without resorting to an exotic matter, has been studied. By using the fact that the Einstein field equations are modified in 3+1 dimensions due to the brane corrections,  we investigate  the exact solutions  which satisfy null energy condition. Asymptotic flatness is an important property of these solutions. We  discuss some   physical and mathematical properties of the solutions. \\
 \textbf{Keywords} : wormholes, brane-world, energy conditions
\end{abstract}
\maketitle
\section{Introduction}
Wormhole are hypothetical objects  in gravitational and cosmology, and a futuristic means to travel in time \cite{Visser}. The term ‘wormhole’ was introduced by Misner and Wheeler  for the first time in 1957 \cite{wheeler}. They have studied the source-free Maxwell equations, coupled to Einstein gravity, with nontrivial topology, to build models for classical electrical charges and all other particle–like entities in classical physics. However, the significant stage of wormhole physics took place through the seminal Morris-Thorne paper \cite{WH}. Although  no observational evidence has been found for the wormhole but many theoretical researchers try to study the theory of wormhole and possible observational consequences. The main ingredient of the wormhole is the violation of null energy condition(NEC). In NEC, $T_{\mu\nu}k^{\mu}k^{\nu}\geq0$, in which $k^{\mu}$ is any null vector and $T_{\mu\nu}$ is the  stress-energy tensor \cite{Visser}. Recent observations approve an accelerated expansion of the cosmos which may be caused by a fluid with  $p=\omega\rho$ with $\omega<-1/3$. The fluid with $\omega<-1$ is called phantom. Since the phantom fluid violates the NEC, it is a good candidate to support wormhole\cite{phantom}. Minimizing the violation of NEC is welcome  in wormhole physics. Some authors have used cut and paste method to construct  wormhole geometries with minimal violation of the NEC \cite{cut}. Building wormholes with a variable equation-of-state (EoS) is  another method to violate the NEC only in a small region in the vicinity of wormhole throat \cite{variable}. In \cite{foad} a class of wormhole solutions with a polynomial EoS has been presented in which the exotic mater is confined to near the wormhole throat. Modified theories of gravity have imposed solutions without violation of the energy conditions. An increasing number of works, in modified  gravity theories, which supports static wormholes both in the vacuum and with matter content  have been presented. Matter content, in some modified gravities, does not violate the NEC by itself. In most of these theories, higher order curvature terms in the right hand side of the Einstein field equations play the role of exotic matter. Static Wormholes in Brans-Dicke \cite{Dicke}, $f(R)$ gravity \cite{tahereh}, curvature matter coupling \cite{modgravity2b}, Einstein-Cartan gravity\cite{Cartan}, and brane-world \cite{a21,23,brane,brane1} have been studied. In the Randall-Sundrum brane-world model, a three-brane is embedded in a five dimensional bulk \cite{a1}. This model was originally introduced to give an alternative to the compactification of the extra dimensions \cite{Brax}. One of the consequences of the Randall–Sundrum brane-world model is  explaining  the hierarchy problem in  particle physics \cite{Brax}. Shiromizu, Maeda and Sasaki have introduced modified Einstein equations on the brane \cite{Shi}.  On the right hand side of modified Einstein field equations, a stress-energy tensor with  higher order terms of $\rho$, $p$ and the term $E_{\mu\nu}$  which come from the projection of 5-dimensional Weyl tensor on  the brane  appear. Bronnikov and Kim have presented  static spherically symmetric Lorentzian wormhole solution in the absence of matter on the brane  \cite{a21}. They have considered $R=0$ where $R$ is the four dimensional Ricci scalar. Lobo has found a class of static solutions in the brane-world scenario with $R\neq0$ \cite{23}. In \cite{brane}, a class of solutions by focusing on the  local conservation equation due to Bianchi identities is investigated.

Broadly speaking, wormholes can be divided into  static and dynamic. Some results show the fact that the properties of dynamic wormholes are quite different from the static ones. Time-dependent spherically symmetric wormholes have been extensively analysed in the literature \cite{Dyn}. Roman has explored a dynamic wormhole immersed in an inflationary background \cite{Roman}. Kim has considered Friedmann-Robertson-Walker (FRW) model with a traversable wormhole \cite{Kim}. Kar and Sahdev have shown that evolving wormholes could satisfy energy conditions \cite{Kar}. Cataldo \textit{et al.}
 have  studied  dynamical wormholes treated in terms of a two fluid system  one with homogeneous and isotropic properties
 and another inhomogeneous and anisotropic \cite{catal}. Their solutions satisfy some energy conditions. Garcia\textit{ et al.} have studied the construction of generic spherically symmetric thin-shell traversable wormhole spacetimes
in standard general relativity (GR) by using the cut-and-paste procedure \cite{Garcia}.
  As it was mentioned, in the modified gravity theory the violation of energy conditions can be avoided due to extra terms  in the field equations. Static wormholes in modified gravity are considered in the literature extensively, limitted work has been done on their dynamic counterparts. In \cite{chak}, the authors have used a reconstruction technique to look for possible evolving wormhole solutions within viable $f(R)$ gravity formalism. Cataldo \textit{et al.} studied the (N + 1) dimensional evolving wormholes supported by a polytropic EoS \cite{catal2}. In \cite{Kord}, higher-dimensional evolving wormholes which satisfy the NEC have been studied. Arellano and Lobo have explored the possibility of evolving  wormhole geometries coupled to nonlinear electrodynamics \cite{electr}. They have shown that  the weak energy condition is satisfied. Mehdizadeh and  Ziaie have investigated the dynamic wormhole solutions in the Einstein-Cartan gravity\cite{Mehdi}. It is a gravitational theory which was used to provide a simple description of the effects of spin on gravitational interactions. The number of papers on  dynamic wormholes in modified gravity is not considerable in contrast to the static models. Wong \textit{et al.} proposed inflating  wormholes in the brane-world  scenario, in which the wormhole is supported by the nonlocal brane-world effects\cite{Wong}. They have explained that a wormhole satisfying some very general initial conditions could turn into a black hole and live forever. In \cite{Wang}, the characteristics and properties of a traversable wormhole constrained by the current astrophysical observations in the framework of the Dvali-Gabadadze-Porrati (DGP) brane-world scenario is investigated.

The present paper investigates the possibility and naturalness of expanding wormholes in the Randall-Sundrum brane-world scenario. We show that dynamical wormholes can exist  in brane-world gravity without the violation of the NEC. The existence  of exact dynamic wormhole solutions in the  brane-world scenario, together with the  weakness and strength of this theory to explain dynamic wormholes are discussed.

 First, we discuss conditions and equations governing wormhole and the  modified equations on the brane. The structure of dynamic wormhole solutions on the brane is very similar to our previous work on static wormhole solutions \cite{brane}. By considering different forms of  $R$, we   present some exact solutions which are asymptotically flat. The paper outline is as follow: In the next section, general properties and equations of the wormhole are presented. In Sec. \ref{sec3} and \ref{sec4}, possible solutions have been studied. The physical and mathematical properties of solutions are investigated in Sec. \ref{sec5}. Concluding remarks are presented in the last section.

\section{Basic formulation of wormhole }
First, we present the basic structure of the wormhole in the brane-world scenario. It should be mentioned that the formalism of dynamic wormholes  is very similar to the static one. We have studied the static wormhole solutions in brane-world scenario in our previous paper \cite{brane}. We use the same formulation for dynamic wormholes. The general spherically symmetric metric of wormhole, in an expanding cosmological background, is as follows
\begin{equation}\label{1}
ds^2=-U(r)dt^2+a(t)(\frac{dr^2}{1-\frac{b(r)}{r}}+r^2(d\theta^2+\sin^2\theta
d\phi^2)).
\end{equation}
where $U(r)> 0$ is called the redshift function and $b(r)$  is called the shape function. Here, $a(t)$ is the cosmological scale factor. The throat of a wormhole connects two universes or distinct parts of the same universe. It is defined by
$r_0$, where
\begin{equation}\label{2}
b(r_0)=r_0.
\end{equation}
Shape function shows the geometrical  form of the wormhole. This function
must satisfy, the so-called flare-out conditions:
\begin{equation}\label{3}
b'(r_0)<1
\end{equation}
and
\begin{equation}\label{4}
b(r)<r,\ \ {\rm for} \ \ r>r_0.
\end{equation}
In order to be asymptotically spatially flat, the metric functions $U(r)$
and $b(r)/r$ should respectively tend to a constant  and zero at $r\rightarrow \infty$.

In the brane-world scenario, our world is a four-dimensional brane which has been embedded in a five-dimensional bulk.  The five-dimensional Einstein equations in the bulk is as follows:
\begin{equation}\label{5}
G_{AB}=-\Lambda g_{AB}+k^{2}T_{AB}.
\end{equation}
Here, Latin indices are related to the bulk and Greek are related to the brane. In \cite{Shi}, authors  have used a covariant formalism to show that Einstein equations on the brane take the following form
\begin{equation}\label{6}
G_{\mu\nu}=-\Lambda g_{\mu\nu}+k^{2}T_{\mu\nu}+\frac{6k^{2}}{\lambda}S_{\mu\nu}-\xi_{\mu\nu}
\end{equation}
where the  gravitational coupling constant and cosmological constant on the brane and bulk are related to each other,
\begin{equation}\label{7}
k^{2}=\frac{\lambda k_{5}^{2}}{6},\quad\quad\Lambda=\frac{1}{2}(\Lambda_{5}+k^{2}\lambda).
\end{equation}
 One should note that $\lambda$ is the tension on the brane and should be considered positive in the Randull Sundrum II model, in order to provide  the correct signature of the gravity. It is clear that there are several correction terms on the right hand side of Eq.(\ref{6}).
The first correction term,
 \begin{equation}\label{8}
S_{\mu\nu}=\frac{1}{12}TT_{\mu\nu}-\frac{1}{4}T_{\mu\alpha}T^{\alpha}_{\nu}+\frac{1}{24}g_{\mu\nu}[3T_{\alpha\beta}T^{\alpha\beta}-T^{2}]
\end{equation}
is a consequence of  the extrinsic curvature terms in the projected Einstein tensor. The last correction term is originated from the nonlocal bulk effect and is defined as $\xi_{\mu\nu}=\delta_{\mu}^{A}\delta_{\nu}^{C}C_{ABCD}n^{B}n^{D}$ which $C_{ABCD}$ is the 5-dimensional Weyl tensor, so  $\xi_{\mu\nu}$ is the projection of Weyl tensor. This  is
  a traceless  tensor. By using this important property from (\ref{6}), one can obtain
 \begin{equation}\label{9}
R=4\Lambda-k^{2}T-\frac{3k^{2}}{2\lambda}(T_{\alpha\beta}T^{\alpha\beta}-\frac{1}{3}T^{2}).
\end{equation}

In  GR theory of gravity, we should consider an anisotropic energy momentum tensor to describe the fluid, which supports the wormhole. But in the brane-world scenario, we can consider an isotropic fluid in the form $T^{\mu}_{\nu}=diag(-\rho,p,p,p)$ where $\rho(r)$ is the energy density and $p(r)$ is the radial pressure. Using this form of energy momentum tensor  gives the components of  $S^{\mu}_{\nu}$ as follow
\begin{eqnarray}\label{10}
S^{t}_{t}&=&\frac{1}{12}\rho^{2}, \nonumber \\
S^r_{r}&=&S^\theta_{\theta}=S^\phi_{\phi}=\frac{1}{12}\rho (\rho+2p).
\end{eqnarray}
Since we have considered an isotropic fluid,  the components of $\xi_{\mu\nu}$ should be in the following form
\begin{eqnarray}\label{12}
\xi^t_{t}&=&\epsilon(r,t),\quad\xi^r_{r}=\sigma_{r}(r,t), \nonumber \\
\xi^\theta_{\theta}&=&\xi^\phi_{\phi}=\sigma_{t}(r,t),
\quad\xi^t_{r}=g^t_{\mu}\xi^{\mu}_{r}=\delta(r,t).
\end{eqnarray}
Bianchi identities and conservation of energy momentum on the brane; $T^{\mu\nu}_{\quad;\nu}=0$ give a relation between the mater on the brane and divergence of $\xi_{\mu\nu}$ as follow
\begin{eqnarray}\label{13}
D^{\mu}\xi_{\mu\nu}&=&\frac{1}{4}k_{5}^{4}[T^{\alpha\beta}(D_{\nu}T_{\alpha\beta}-D_{\beta}T_{\nu\alpha}) \nonumber \\
&+&\frac{1}{3}(T_{\mu\nu}-g_{\mu\nu}T)D^{\mu}T].
\end{eqnarray}
Using the line element (\ref{1}) and $T^{\mu\nu}_{\quad;\nu}=0$ leads to
\begin{equation}\label{13a}
U'(\rho+p)=-2p'U
\end{equation}
and
\begin{equation}\label{13b}
\dot{\rho}\,a=-3\,\dot{a}(\rho+p).
\end{equation}
Note, overdots denote derivatives with respect to $t$ and the
primes are derivatives with respect to $r$ .
 Ricci scalar, corresponds to  metric  (\ref{1}), is
\begin{eqnarray}\label{14}
R&=&-\frac{1}{2r^{2}a^{2}U^{2}}[4U'rU-U'rUb'+2U''r^{2}U
\nonumber \\ &-&2U''rUb
-U'^{2}r^{2}+U'^{2}rb-3U'Ub
\nonumber \\&-&4U^{2}b'
-12\ddot{a}ar^{2}U-12\dot{a}^{2}r^{2}U],
\end{eqnarray}
which can be presented in the following form
\begin{equation}\label{n14}
R=-\frac{1}{2r^{2}a^{2}U^{2}}[F(r,U,b)-12r^{2}U(\ddot{a}a+\dot{a}^{2})].
\end{equation}
Here,
\begin{eqnarray}\label{n141}
F(r,U,b)=4U'rU-U'rUb'+2U''r^{2}U \nonumber \\
-2U''rUb-U'^{2}r^{2}+U'^{2}rb-3U'Ub-4U^{2}b'.
\end{eqnarray}
Another form of Ricci scalar in terms of energy density is given by
\begin{equation}\label{18}
R=k^{2}(\rho-3p)-\frac{3k^{2}}{2\lambda}[\rho^{2}+3p^{2}-\frac{1}{3}(\rho-3p)^{2}].
\end{equation}
This equation can be presented in the form
\begin{equation}\label{18a}
p=\frac{1}{3}[\rho\frac{1-\frac{\rho}{\lambda}}{1+\frac{\rho}{\lambda}}-\frac{R}{k^2(1+\frac{\rho}{\lambda})}]
\end{equation}
which can be interpreted as an EoS for matter supporting wormhole.

Now, we want to present  a mathematical full analysis about the possible method for constructing wormholes with different forms of Ricci scalar. We have six unknown functions $\epsilon(r,t), \sigma_{r}(r,t), \sigma_{t}(r,t), \rho(r,t)$, $p(r,t)$,$\delta(r,t)$ and three unknown functions in the metric $b(r)$,$U(r)$, $a(t)$. On the other hand, we have eight equations, four of which come from Einstein equations, two from conservation of energy and two from the constraint on $R$ and Eq.(\ref{9}) which is a consequence of the traceless property of $\xi_{\mu\nu}$. In our algorithm, we use the constraint on $R$  and Eqs.(\ref{13a}) and (\ref{13b}) to find $a(t)$, $\rho$ and $p$. Since the number of total equations is one less than unknown functions,  we consider $b(r)$ or $U(r)$ arbitrary and then try to find the other unknown functions. It should be noted that other algorithms can be used instead of considering a known shape or redshift function but the presented algorithm seems to be less complicated  in contrast to the other ones.

 After using Eqs. (\ref{6},\ref{10},\ref{12}), from Einstein equations on the brane, we have
\begin{equation}\label{18aa}
\epsilon(r,t)= k^2[\rho(1+\frac{\rho}{2\lambda})]+G^t_t
\end{equation}
\begin{equation}\label{18b}
\sigma_{r}(r,t)=k^2[p(1+\frac{\rho}{\lambda})+\frac{\rho^2}{2\lambda}]+G^r_r
\end{equation}
\begin{equation}\label{18c}
\sigma_{t}(r,t)=k^2[p(1+\frac{\rho}{\lambda})+\frac{\rho^2}{2\lambda}]+G^\theta_\theta
\end{equation}
and
\begin{equation}\label{18d}
G^t_r= g^t_\mu G^\mu_r=\delta .
\end{equation}

 In the next section, we use Eq.(\ref{18a}) with different forms of $R$ to find wormhole solutions. In our strategy, we put Eq.(\ref{18a}) in (\ref{13b}) and then integrate to find a solution for $\rho$. The cosmological scale factor can be determined from Eq.(\ref{14}) and condition on Ricci scalar. In the next step, we consider a known shape or redshift function and try to find the other by using Eq.(\ref{n14}).
After finding shape and redshift functions,  metric is completely determined so we should try to find $\xi_{\mu\nu}$ and  $T_{\mu\nu}$  through Eqs. (\ref{9}) - (\ref{18d}). It is very complicated to work with a general shape or redshift  function so in the next section, we chose a specific shape or redshift function  and then discuss the solutions for $\xi_{\mu\nu}$ and  $T_{\mu\nu}$.

\section{Solutions with constant redshift function }\label{sec3}
In this section, we try to find wormhole solutions with constant redshift function. Constant redshift function guaranties the absence of horizon around the throat. If we set $U(r)=1$ then Eqs.(\ref{13a}) and (\ref{13b}) imply that $\rho$ and $p$ are only functions of $t$. So the Ricci scalar should also be a function of $t$. Now, Eq.(\ref{n14})can be separated into radial and temporal equations
\begin{eqnarray}\label{16a}
F(r,U,b)=-4b'=C0\,r^2
\end{eqnarray}
 and
 \begin{equation}\label{16}
12(\ddot{a}a+\dot{a}^{2})= a^2 R(t)-C_0.
\end{equation}
Equation (\ref{16a}) leads to
 \begin{equation}\label{16b}
b(r)=C_0 r^3+b_0.
\end{equation}
which shows that the only asymptotically flat solution is related to $C_0=0$ and $b_0=r_0$ ($C_0\neq 0$ corresponds to ds/Ads asymptotic). This solution is called spatial-Schwarzschild. From now, we study the possibilities of solutions with different forms of Ricci scalar.

\subsection{Wormholes with  $R=0$}\label{subsec1}
Solutions with $R=0$ is of great importance in GR. Dadhich \textit{et al.}  have studied a class of static  wormhole solutions with $R=0$ \cite{R}. Now, we discus the dynamic wormhole solutions with $R=0$ and $U(r)=1$ in the brane-world scenario. We can put $R=0$ in (\ref{18a}) then
\begin{equation}\label{20}
p=\frac{\rho}{3}\frac{1-\frac{\rho}{\lambda}}{1+\frac{\rho}{\lambda}}.
\end{equation}
Taking into account Eq.(\ref{13b}) yields
\begin{equation}\label{20a}
\ln(c\,a) =-\int\frac{d\rho}{3(\rho+p)}
\end{equation}
where $c$ is a constant of integration. By using (\ref{20}) and (\ref{20a}), one can find
\begin{equation}\label{20b}
\rho_{\pm}(t)=-\lambda\pm\sqrt{\lambda^2+\frac{c}{a(t)^4}}.
\end{equation}
From the constraint $R=0$ and Eq. (\ref{16}), we find
\begin{equation}\label{20c}
a(t) =\sqrt{c_1 t+c_2 }
\end{equation}
where $c$, $c_1$ and $c_2$ are constants of integration. Some physical and mathematical properties of this kind of solutions will be studied in Sec. \ref{sec5}.

\subsection{Wormholes with constant Ricci scalar }\label{subsec2}
Considering $R=cte$ yields
\begin{equation}\label{m}
R\,a^{2} -6\frac{(\ddot{a}a+\dot{a}^{2})}{U^2}=-\frac{1}{2r^{2}U^{2}}F(r,U,b).
\end{equation}
This equation is acceptable when $U(r)$ is a constant. So solutions with constant Ricci scalar are possible only for constant redshift function.
We can use the same algorithm to find exact wormhole solutions for constant Ricci scalar. Using (\ref{20a}) leads to
\begin{equation}\label{20d}
\rho_{\pm}(t)=-\lambda \pm\sqrt{\lambda^2 + \frac{R}{2}\,\lambda+\frac{c_3}{a(t)^4}}
\end{equation}
where $R$ and $c_3$ are constants. Also we use $R=cte$ in (\ref{16}) to find the scale factor as follows
\begin{equation}\label{20e}
a(t) = (c_4 e^{\sqrt{\frac{R}{6}}t}+c_5 e^{-\sqrt{\frac{R}{6}}t})^{1/2},
\end{equation}
where $c_4$ and $c_5$ are constants of integration.

\subsection{$R$ as a function of time}\label{subsec3}
The Ricci scalar is only a function of time in standard cosmological models. Now, we try to find wormhole solutions when $R$ is time-dependent. So we set
\begin{equation}\label{20f}
R(t) =12\frac{f(t)}{a^2}
\end{equation}
where $f(t)$ is an arbitrary function of time. Using this assumption for $R$, give the following general solution for scale factor,
 \begin{equation}\label{20g}
a(t) =\sqrt{2(t\int f(t)\, dt -\int tf(t)\, dt +c_6\,t+c_7)}.
\end{equation}
As an example for $f(t)=B\,t^n$ where $n\neq-2$ and $n\neq-1$, one can find
\begin{equation}\label{20h}
a(t) =\sqrt{2(B\frac{t^{n+2}}{n^2+3n+2} +c_6\,t+c_7)}.
\end{equation}
Finding solutions for $\rho(t)$ when $f(t)$ is a general function is very difficult so we investigate two special cases.
 First, we consider $f(t)=B$ which corresponds to $n=0$. In this case, scale factor and $\rho(t)$ are as follow
\begin{eqnarray}\label{20i}
a(t)=\sqrt{Bt^2+2c_6t+2c_7} \nonumber, \\
 \rho_{\pm}(t)=-\lambda\pm \sqrt{\lambda^2+\frac{c_8}{a(t)^2}}.
  \end{eqnarray}
The solutions, related to $f(t)=\frac{B}{t^2}$ are
\begin{eqnarray}\label{20j}
a(t)=\sqrt{\frac{B}{3t^2}+2c_6t+2c_7} \nonumber \\
 \rho_{\pm}(t)=-\lambda\pm \sqrt{\lambda^2+\frac{g(t)}{a(t)^2}}
  \end{eqnarray}
where
\begin{equation}\label{20k}
g(t)=c_8-8c_9t^{-3}+\frac{4}{3}Bt^{-6}
\end{equation}
In Sec. \ref{sec5}, we will analyze the  physical properties of these solutions.

\subsection{Vacuum and constant solutions }\label{subsec4}
Static wormhole with vanishing energy density and pressure has been studied in the brane-world scenario \cite{a21,23}. These  are the first presented class of solutions in this category. Let us investigate the vacuum solutions in the dynamic case. Vacuum solutions satisfy Eqs. (\ref{13a}) and (\ref{13b}). Putting $p=\rho=0$ in (\ref{18}) gives $R=0$. So vacuum solution is a special case of vanishing Ricci scalar. For constant redshift function, the vacuum solution  corresponds to  $c=0$ in Eq. (\ref{20b}) which gives $p=\rho=0$ and $p=-\rho=2\lambda$. The former is desired vacuum solution for constant redshift function.

 Another interesting case is solutions with constant energy density. If we consider $\rho=cte$ then Eq.(\ref{13b}) implies that $a(t)=cte$ or $\rho=-p$. The former is not acceptable according to recent observations but the latter has the necessary condition to find solutions with constant energy density. It is obvious that constant energy density leads to vanishing or constant Ricci scalar. First, we  seek vanishing Ricci scalar solutions with $p=-\rho$. Putting $p=-\rho$ and $R=0$ in Eq.(\ref{18}) yields
\begin{equation}\label{200}
\rho_{\pm} =-\lambda \pm\sqrt{\lambda^2}
\end{equation}
which is the same as the solution (\ref{20b}) with vanishing constant of integration ($c=0$). The same analyses for $R=cte$ and $p=-\rho$ give
\begin{equation}\label{201}
\rho_{\pm} =-\lambda \pm\sqrt{\lambda^2+\frac{R\lambda}{2}}
\end{equation}
 which is the special case of the solution (\ref{20d}) with vanishing constant of integration ($c_3=0$). From the form of Eq.(\ref{201}),  one can deduce that $R>-2\lambda$.
 If we consider $R=\alpha \lambda$ then it is clear that
 \begin{equation}\label{201a}
\rho_{\pm} =(-1\pm \sqrt{1+\frac{\alpha}{2}})\lambda.
\end{equation}
 It is obvious that $\alpha\geq-2$. The other form of vacuum solution is related to nonconstant redshift function, which is considered in the next section.

\section{Nonconstant redshift function }\label{sec4}
In GR, dynamic wormholes with nonconstant redshift function are not investigated in the literature, since a nonconstant redshift function produces off-diagonal elements in the Einstein tensor. Brane-world  gives us the opportunity to investigate wormhole solutions with nonconstant redshift function. In this scenario, the off-diagonal elements in Einstein tensor could be interpreted as a consequence of Weyl tensor. Let us study the possibility of wormhole solutions with nonconstant redshift function. We consider several possibilities for $R$ and then try to find solutions. Our motivation, to consider these possibilities for Ricci scalar, is based on the fortunate use of the separation of variable method in the resulting equations.

\subsection{R=0 solutions }\label{subsec5}

Considering $R=0$ in Eq.(\ref{n14}) and using the separation of variables gives
\begin{equation}\label{01}
F(r,U,b)=C_0\,r^2
\end{equation}
 and
 \begin{equation}\label{02}
12(\ddot{a}a+\dot{a}^{2})= C_0.
\end{equation}
One can see that Eq.(\ref{02}) is the same as Eq.(\ref{16}) for $R=0$. Solutions for Eq.(\ref{01}) when $C_0=0$ are the same as solutions with $R=0$ in static case. The general solution for Eq.(\ref{02}) is
\begin{equation}\label{02a}
a(t) =\sqrt{C_0 t^2+c_1 t+c_2 }.
\end{equation}
For the static wormhole, the energy density is as follows \cite{brane}
 \begin{equation}\label{03}
\rho_{\pm}=-\lambda\pm\sqrt{\lambda^2+\gamma(r)^2},
\end{equation}
where
\begin{equation}\label{04}
\gamma(r)=\frac{-U(r)\pm\sqrt{U(r)^2-4C^2\lambda^2}}{2C}.
\end{equation}
Here, $C$ is an integration constant. This solution is based on Eq.(\ref{13a}). Since Eq.(\ref{13a}) must be satisfied in the dynamic case, the dynamic solutions should be consistent with this solution.  Because of the aforementioned reasons, $C$ should be considered  as a function of $t$ in the dynamic case. In the dynamic case, one can find solutions for $\rho$ which are based on Eq.(\ref{13b}). As it was mentioned in Sec. \ref{subsec1} these solutions are in the form of Eq.(\ref{20b})
where $c$ is a constant of integration. So, for consistency between these solutions, leads to
\begin{equation}\label{05}
\gamma(r,t)=\frac{-U(r)\pm\sqrt{U(r)^2-4C(t)^2\lambda^2}}{2C(t)}=\frac{\sqrt{c(r)}}{a(t)^2}.
\end{equation}
But the form of these solutions do not allow us to maintain this consistency. So one can deduce that there are not any temporal or radial solutions in this case. It is interesting to note that constant energy density or vacuum solutions satisfy Eqs.(\ref{13a}) and (\ref{13b}), simultaneously. Therefore, the only possible solutions are constant energy density, $p=-\rho=2\lambda$, and vacuum solution  with non-constant redshift function and $R=0$. One can construct a general class of these solutions with different forms of shape and redshift functions. In fact, one should consider a known shape function and try to find $U(r)$ from Eq.(\ref{01}) or considering a known redshift function and try to find $b(r)$.
 The simplest case for shape function is $b(r)=r_{0}$. Solving Eq.(\ref{01}) for a general $C_0$ is very difficult so, we put $C_0=0$.  This particular choice for $b(r)$ and $C_0$ leads to
\begin{equation}\label{22a}
U(r)=\left[C_{1}+C_{2}\sqrt{(1-\frac{r_{0}}{r})}\right]^{2} .
\end{equation}
Here, $C_{1}$ and $C_{2}$ are constants of integration. This solution is related to the one which has been analysed in detail in the context of ordinary GR in ref \cite{R}. So, we can conclude that solutions which have been presented in  \cite{R} can be considered as compatible solutions for dynamic wormhole solution in brane-world scenario with constant ($p=-\rho=2\lambda$) or vacuum energy density. The related cosmological scale factor is $a(t) =\sqrt{c_1 t+c_2 }$ which is similar to radiation FRW flat model. Let us discuss a special solution in detail.
If we put $C_{2}=0$ then
 \begin{equation}\label{24a}
U(r)=C_{1}^{2} .
\end{equation}
which is the case related to constant redshift function. In this case, $C_1$ can be set equal to unity. As another case, by choosing $C_{1}=C_{2}=\frac{1}{2}$ , $U(r)$ reduces to
\begin{equation}\label{25a}
U(r)=\left[\frac{1}{2}+\frac{1}{2}\sqrt{(1-\frac{r_{0}}{r})}\right]^{2} .
\end{equation}
Now, these two forms of $U(r)$  will provide completely different results for other components. In the first case, non vanishing components of Einstein tensor are as follow
 \begin{eqnarray}\label{26a}
G^{t}_{t}=\frac{-3}{4t^{2}},\quad G^{r}_{r}=\frac{1}{4t^{2}}-\frac{r_{0}}{r^{3}t},\quad\\
\nonumber G^{\theta}_{\theta}=G^{\phi}_{\phi}=\frac{1}{4t^{2}}+\frac{r_{0}}{2r^{3}t}
  \end{eqnarray}
in which no off-diagonal element appears. So, in this case
   \begin{equation}\label{27a}
G^t_{r}=\delta=\xi^t_{r}=0.
\end{equation}

After using equation (\ref{6}) and considering $p=-\rho=2\lambda$, we have
\begin{equation}\label{30a}
\epsilon(r,t)= \frac{-3}{4t^{2}}=-f(r,t),
\end{equation}
\begin{equation}\label{31a}
\sigma_{r}(r,t)=\frac{3}{4t^{2}}-\frac{3 r_{0}}{r^{3}t}=f(r,t)-2 g(r,t),
\end{equation}
\begin{equation}\label{32a}
\sigma_{t}(r,t)= \frac{3}{4t^{2}}+\frac{3 r_{0}}{2r^{3}t}=f(r,t)+g(r,t),
\end{equation}
where
\begin{equation}\label{33a}
f(r,t)= \frac{3}{4t^{2}}, \qquad g(r,t)=\frac{3 r_{0}}{2r^{3}t} .
\end{equation}

In the second case, we can follow the same calculation which leads to
\begin{equation}\label{37a}
\delta(r,t)=G^t_{r}=\frac{f_1(r,t)}{g_1(r,t)},
\end{equation}
and
\begin{eqnarray}\label{38a}
\epsilon(r,t)=\frac{f_2(r,t)}{g_1(r,t)},\nonumber \\
\sigma_{r}(r,t)=\frac{f_3(r,t)}{g_1(r,t)}, \nonumber \\
 \sigma_{t}(r,t)=\frac{f_4(r,t)}{g_1(r,t)}.
\end{eqnarray}
$f_i(r,t)$ where $i=1..4$ and $g_1(r,t)$ are as follow
\begin{eqnarray}\label{39a}
f_1(r,t)&=& -\frac{1}{2}(rr_0^2-r^2r_0) ,\nonumber \\
f_2(r,t)&=& -3r^4(1-\frac{r_0}{r})^{1/2} ,\nonumber \\
f_3(r,t)&=& -\frac{1}{4(1-\frac{r_0}{r})}(-8r^2_0t-r_0^2t(1-\frac{r_0}{r})^{1/2} \nonumber \\
&+&\frac{4r_0^3t}{r}+\frac{r_0^3t}{r}(1-\frac{r_0}{r})^{1/2} +4tr_0r\nonumber \\
&+& 5r^2_0t(1-\frac{r_0}{r})^{3/2}+4r_0tr(1-\frac{r_0}{r})^{5/2} \nonumber \\
&-& 4r^4(1-\frac{r_0}{r})^{3/2}),\nonumber \\
f_4(r,t)&=& \frac{1}{2}(r_0tr-r_0^2t+tr^2_0(1-\frac{r_0}{r})^{1/2} \nonumber \\
&+&tr_0r(1-\frac{r_0}{r})^{3/2}+2r^4(1-\frac{r_0}{r})^{1/2}),\nonumber \\
 g_1(r,t)&=&r^4t^2(1+ (1-\frac{r_0}{r})^{1/2}))^2(1-\frac{r_0}{r})^{1/2}.
\end{eqnarray}

By choosing particular choices for $U(r)$, we can find shape function through Eq.(\ref{6}). For example, redshift function, is given by
\begin{equation}\label{22}
U(r)=1+\frac{r_0}{r},
\end{equation}
has a maximum at the throat and tends to unity at large distances from the throat and seems to be a good candidate.
Putting this function into Eq.(\ref{18}) leads to
 \begin{equation}\label{23}
b(r)=(\frac{\alpha_{1}(r +r_0)}{4r+3r_0}+\frac{r_0^2}{(4r+3r_0)}) .
\end{equation}
in which $\alpha_{1}$ is a constant of integration. We should check that this shape function satisfies all of the necessary conditions to present a Lorentzian wormhole. First, we use the condition $b(r_0)=r_0$ to  find $\alpha_{1}=3r_0$. It is obvious that flare out condition is satisfied. Next, we
check  condition (\ref{4}) by defining
\begin{equation}\label{25}
\psi(x) \equiv\frac{ b(r)-r}{r_0} =\frac{3x+4}{4x+3}-x .
\end{equation}
where $x=\frac{r}{r_o}$.
By imposing condition (\ref{4}) we should have
 \begin{equation}\label{26}
\psi(x) <0 .
\end{equation}
\begin{figure}
\centering
  \includegraphics[width=3.2in]{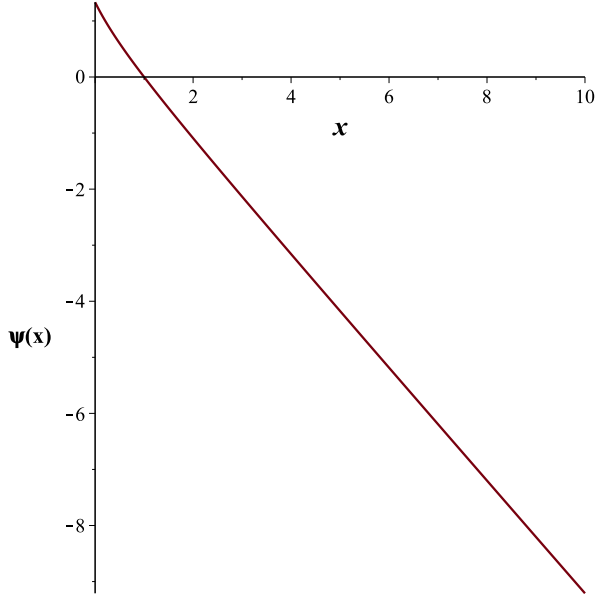}
\caption{The plot depicts the function $\psi(x)$ against $x$. It shows that $\psi(x)<0$ thorough the entire range of $1<x$  which means the $b(r)<r$ is satisfied everywhere.
 See the text for details.}
 \label{fig1}
\end{figure}

We have plotted $\psi(x)$ against $x$ in Fig.(1). This figure indicates that condition (\ref{4}) is satisfied. So the shape function (\ref{23}) is a suitable one. For the sake of simplicity,  we set $r_0=1$ in the recent part of this paper. By using this special solution  for $b(r)$, we can find the term related to nonlocal bulk effect. In this case, all of the wormhole conditions discussed in the previous section are satisfied, and the metric  is as follow :
\begin{equation}\label{27}
ds^2=-(1+\frac{1}{r})dt^2+\frac{tdr^2}{1-\frac{3r+4}{r(4r+3)}}+tr^2(d\theta^2+\sin^2\theta
d\phi^2).
\end{equation}
 For this metric, non-vanishing components of the Einstein tensor are as follow
 \begin{eqnarray}\label{28}
G^{t}_{t}&=&\epsilon(r,t)=-\frac{1}{4}\frac{48r^5+72r^4-27r^3-28tr-28t}{r^2t^2(r+1)(4r+3)^2},  \nonumber \\
G^{r}_{r}&=&\sigma_r(r,t)=\frac{1}{4}\frac{4r^4+3r^3-28tr-28t}{r^2t^2(r+1)(4r+3)}, \nonumber \\
G^{\theta}_{\theta}&=&G^{\phi}_{\phi}=\sigma_t(r,t)  \nonumber \\
&=&\frac{1}{4}\frac{16r^5+24r^4+9r^3+56tr^2+84tr+28t}{r^2t^2(r+1)(4r+3)^2}
  \end{eqnarray}
  and the off-diagonal component reads
 \begin{equation}\label{29}
  G^{t}_{r}=\delta(r,t)=\frac{1}{2t(r+1)^2}.
\end{equation}

\begin{figure}
\centering
  \includegraphics[width=3.2in]{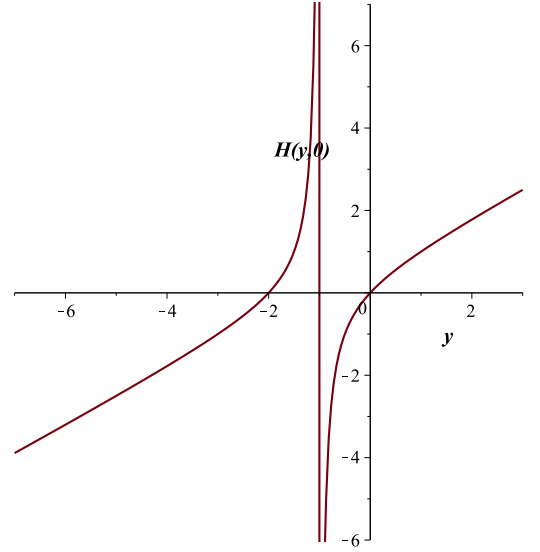}
\caption{The plot depicts the function $H(y,0)$ against $y$. It is clear that $H(y,0)$ is positive thorough the entire ranges of $-2<y<-1$ and $0<y$ which means the NEC is not violated in these ranges.
 See the text for details.}
 \label{fig2}
\end{figure}

\section{Solutions properties }\label{sec5}
In this section, we study some physical properties of the solutions.
First, let us have a look at the null energy condition; for the nonconstant redshift function, the only possible solutions are vacuum or $p=-\rho$ which satisfies NEC. Now, we try to investigate constant redshift function solutions.
From Eq.(\ref{18a}), NEC can be checked using the function
\begin{equation}\label{40a}
 H(y,R)=\rho+p=\lambda\,y(1+\frac{1}{3}\frac{1-y}{1+y})-\frac{R}{3(1+y)}.
\end{equation}
where
\begin{equation}\label{40aa}
 y(t)=\frac{\rho(t)}{\lambda}.
\end{equation}
In order to check NEC, we should analyze $H(y,R)$ for different forms of $R$. In the first case, solutions with $R=0$ are analysed. For this class of solutions, we have plotted $H(y,0)$ as a function of $y$ in Fig. \ref{fig2}. It is clear that NEC is satisfied through the range $-2\leq y\leq -1$ and $0\leq y$. So, we should investigate the possible case of $y$. Equation (\ref{20b}) implies that the form of $y$ depends on  constant of integration and plus/mines sign in these solutions. In general, one can say that in the limit $t\rightarrow 0$,
\begin{equation}\label{40b}
 y_{\pm}(0)=\frac{\rho_{\pm}^0}{\lambda}=\lim_{t\rightarrow 0}y(t)=-1\pm\sqrt{1+\frac{c}{c_2}}
\end{equation}
and in the limit $t\rightarrow \infty$
\begin{equation}\label{40c}
 y_{\pm}(\infty)=\frac{\rho(\infty)}{\lambda}=\lim_{t\rightarrow 0}y(t)=-1\pm\sqrt{1}
\end{equation}
\begin{figure}
\centering
  \includegraphics[width=3.2in]{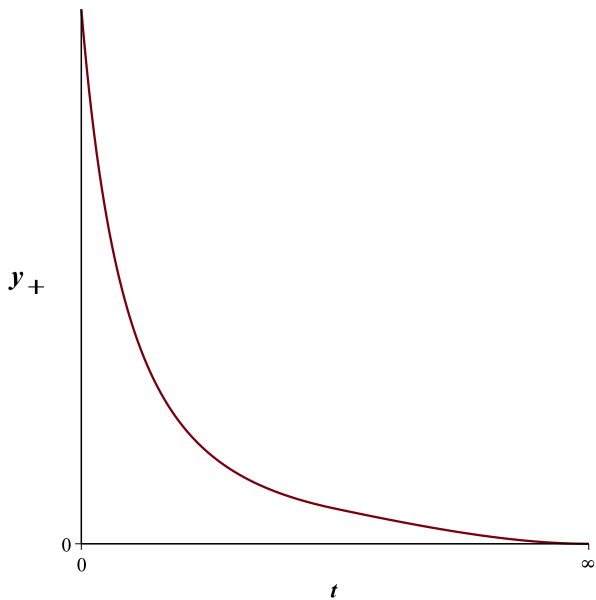}
\caption{The plot depicts the general behavior of  $y_{+}(t)$ against $t$. It is clear that $y_{+}(t)$ has a maximum at $t=0$ and tends to zero as $t\rightarrow \infty$.
 See the text for details.}
 \label{fig3}
\end{figure}
We have plotted the general behaviour of $y_{\pm}(t)$ as a function of $t$ in Figs. \ref{fig3} and \ref{fig4} . From these figures, one can deduce that positive energy density has a maximum at $t=0$ and tends to zero when time elapses sufficiently due to the expansion of the Universe. The negative energy density has a minimum  but it tends to $-2\lambda$ instead of vanishing energy density. The maximum absolute value  of negative and positive $\rho$ are absolute value  of $\rho_{-}(0)$ and $ \rho_{+}(\infty)$ respectively. From Figs.\ref{fig2}-\ref{fig4},  we can conclude that solutions with $\rho_+$ seem to be a better candidate for satisfying energy conditions. These solutions also provide a vanishing $\rho$ at large time which seems to be more physical. For constant redshift function, the violation of NEC is dependent on $R$. We have plotted $H(y,R)$ for some different values of $R$ in Fig.5. If we consider only the positive solutions, then the behavior of this kind of solutions is the same as vanishing Ricci scalar. But the interval of violation of the NEC may be a little different.

\begin{figure}
\centering
  \includegraphics[width=3 in]{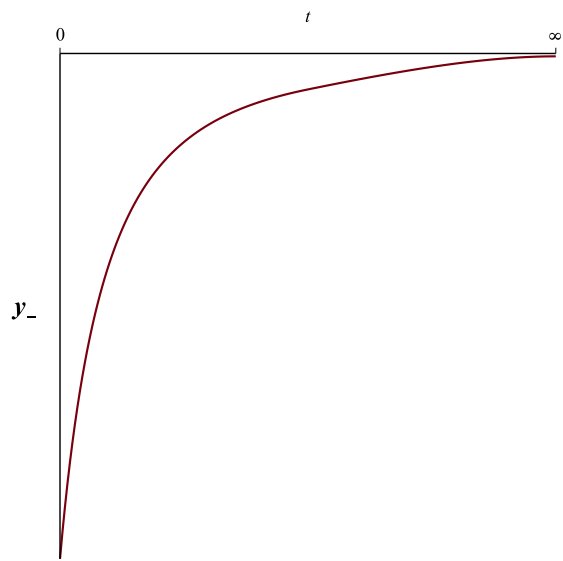}
\caption{The plot depicts the general behavior of  $y_{-}(t)$ against $t$. It should be noted that $y_{-}(t)$ has a minimum at $t=0$ and tends to $-2$ as $t\rightarrow \infty$.
 See the text for details.}
 \label{fig4}
\end{figure}

\begin{figure}
\centering
  \includegraphics[width=3 in]{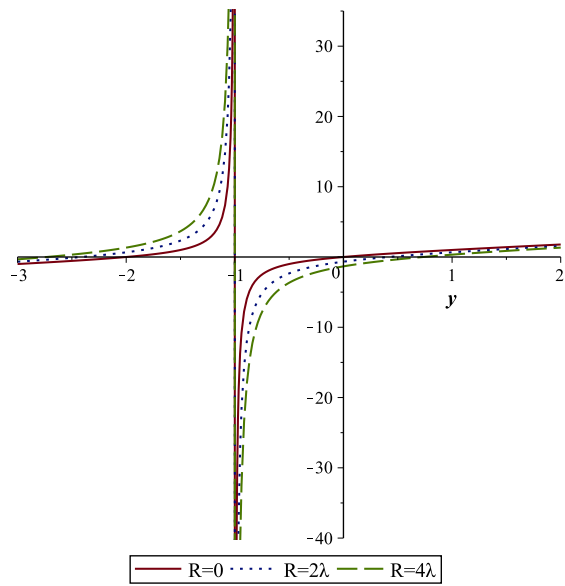}
\caption{The plot depicts the general behavior of $H(y,0)$ (solid line),$H(y,2\lambda)$ (dotted
line) and $H(y,4\lambda)$ (dashed line), against $y$. It is clear that the general behaviour of the $H(y,R)$ is the same for different values of $R$. The only difference is that the range of $y$ for the violation of NEC has been changed.
 See the text for details.}
 \label{fig5}
\end{figure}

Now, let us have a look at the EoS for solutions. As it was mentioned before, the general form of EoS can be described by Eq.(\ref{18a}).But there are some forms of $R$ and the redshift function, this equation takes a simpler form. It has been shown that $p=-\rho$ is the linear form of Eq.(\ref{18a}) for non constant solutions with $\omega=-1$. For constant redshift function solutions, the general behaviour of EoS is not linear but it can be shown that the asymptotic behavior is linear. To show this point, we have defined the effective EoS parameter as follow
 \begin{equation}\label{41}
 \omega_{eff}(r,t)=\frac{p(r,t)}{\rho(r,t)}.
\end{equation}
The asymptotic behavior of this function presents the asymptotic behavior of EoS.
As the first example, the asymptotic behaviour of solution which has been presented in Sec. \ref{subsec1} ($R=0$ and $U=1$) is as follow
 \begin{eqnarray}\label{42}
 \rho_{+}(\infty)&=&\lim_{t\rightarrow \infty }\rho_{+}(t)\longrightarrow 0 \nonumber \\
 p_{+}(\infty)&=&\lim_{t\rightarrow \infty }p_{+}(t)\longrightarrow 0 \nonumber \\
 \omega_{eff}(\infty)&=&\lim_{t\rightarrow \infty}\frac{p(t)}{\rho(t)}\longrightarrow \frac{1}{3}.
\end{eqnarray}
and
\begin{eqnarray}\label{43}
 \rho_{-}(\infty)&=&\lim_{t\rightarrow \infty }\rho_{-}(t)\longrightarrow -2\lambda \nonumber \\
 p_{-}(\infty)&=&\lim_{t\rightarrow \infty }p_{-}(t)\longrightarrow 2 \lambda \nonumber \\
 \omega_{eff}(\infty)&=&\lim_{t\rightarrow \infty}\frac{p(t)}{\rho(t)}\longrightarrow -1.
\end{eqnarray}
These results show that the EoS has a linear form asymptotically. Also it is obvious that the solution with $+$ sign has more consistency with recent observational data.
For solution with $U=1$ and $R=cte$ if we consider $R=\alpha \lambda$ then one can find
\begin{eqnarray}\label{44}
 \rho_{\pm}(\infty)&=&\lim_{t\rightarrow \infty }\rho_{\pm}(t)\longrightarrow (-1\pm\sqrt{1+\frac{\alpha}{2}})\lambda=n_{\pm}\lambda \nonumber \\
 p_{\pm}(\infty)&=&\lim_{t\rightarrow \infty }p_{\pm}(t)\longrightarrow \frac{1}{3}n_{\pm}(\frac{1-n_{\pm}-\frac{\alpha}{n_{\pm}}}{1+n_{\pm}}) \lambda \nonumber \\
 \omega_{eff \pm}(\infty)&=&\lim_{t\rightarrow \infty}\frac{p_{\pm}(t)}{\rho_{\pm}(t)}\longrightarrow \frac{1}{3}(\frac{1-n_{\pm}-\frac{\alpha}{n_{\pm}}}{1+n_{\pm}}).
\end{eqnarray}
\begin{figure}
\centering
  \includegraphics[width=3 in]{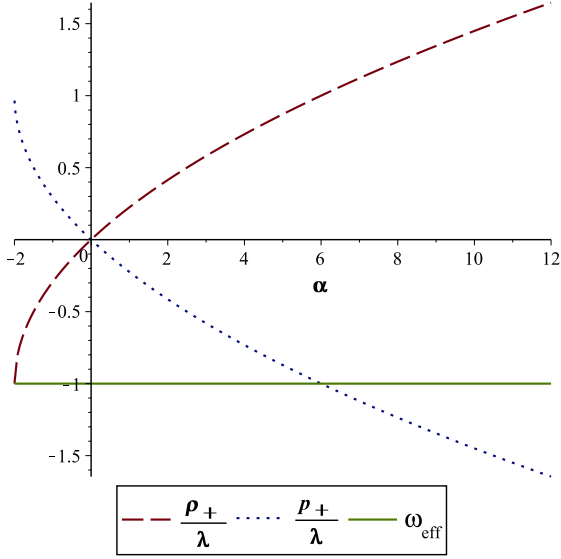}
\caption{The plot depicts the general behavior of $\frac{\rho_+}{\lambda}$ (dashed line), $\frac{p_+}{\lambda}$ (dotted
line) and $\omega_{eff}$ (solid line), against $\alpha$. It is clear that $\rho_+$ is positive through the range $\alpha \geq 0$ and changes the sign in $\alpha=0$. It shows that the range $\alpha \geq 0$ is more physically.
 See the text for details.}
 \label{fig6}
\end{figure}
 One should consider that $\alpha \geq -2$ is the acceptable range in these equations. It is easy to show that
  $\omega_{eff \pm}(\infty)=-1$. This is an important result which confirms that  EoS has an asymptotically  linear form. We have plotted $p(\infty)$, $\rho(\infty)$ and $\omega_{eff}(\infty)$ as a function of $\alpha$ in Figs. \ref{fig6}  and \ref{fig7} respectively for $+$ and $-$ signs. Figure \ref{fig6} implies that $\rho_+$ is positive through the range $\alpha \geq 0$ so solutions with $R\geq 0$ are more physical. On the other hand, Fig. \ref{fig7} indicates that $\rho_-$ is negative in the entire range of $\alpha$. This kind of solution is not  interesting in the brane-world scenario.
   It is difficult to study  EoS of solutions with a time-dependent $R$ in a general form. So we study the special cases which has been presented in Sec. \ref{subsec3}. The asymptotic behaviour of solution (\ref{20i})  is as follow
    \begin{eqnarray}\label{45}
 \rho_{+}(\infty)&=&\lim_{t\rightarrow \infty }\rho_{+}(t)\longrightarrow 0 \nonumber \\
 p_{+}(\infty)&=&\lim_{t\rightarrow \infty }p_{+}(t)\longrightarrow 0 \nonumber \\
 \omega_{eff}(\infty)&=&\lim_{t\rightarrow \infty}\frac{p_{+}(t)}{\rho_{+}(t)}\longrightarrow \frac{1}{3}(1-\frac{24}{c_8})\nonumber \\
 R(\infty)&=&\lim_{t\rightarrow \infty }R(t)\longrightarrow 0 .
\end{eqnarray}
This indicates that the general asymptotic behaviour of this kind of solutions is the same as the solutions with vanishing Ricci scalar. The only difference is the $\omega_{eff}(\infty)$ which is dependent, for time-dependent solutions,  on constant of integration $c_8$. To summarize, solutions with time-dependent $R$  and $U=1$ seem to be better candidates in contrast to the other form of Ricci scalar to present wormhole solutions in the brane-world scenario.

\begin{figure}
\centering
  \includegraphics[width=3 in]{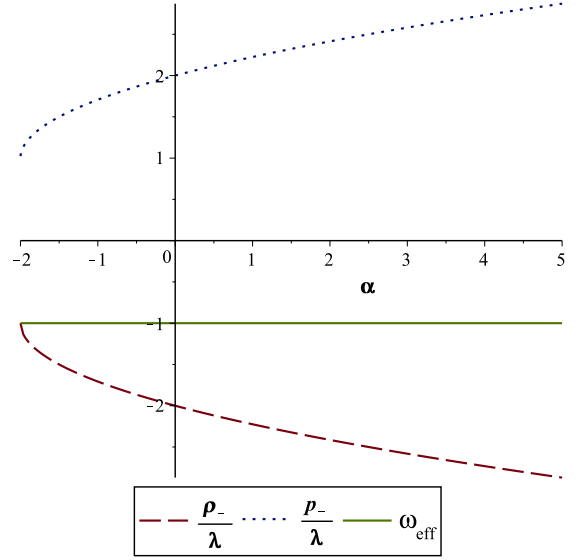}
\caption{The plot depicts the general behavior of $\frac{\rho_-}{\lambda}$ (dashed line), $\frac{p_-}{\lambda}$ (dotted
line) and $\omega_{eff}$ (solid line), against $\alpha$. It is clear that $\rho_-$ is negative through the entire range $-2\leq \alpha$ but $p_-$ is positive. It shows that this class of solutions is not of great interest.
 See the text for details.}
 \label{fig7}
\end{figure}

\section{Concluding remarks}
In the present paper, we have used Randall-Sundrum brane-world  scenario to  find asymptotically flat (FRW) wormhole solutions in a cosmological background with different forms of Ricci scalar. Selecting different forms of Ricci scalar and using the separation of variables method lead to two distinct temporal and radial  equations. Our solutions are based on conservation of energy which leads to Eqs.(\ref{13a}) and (\ref{13b}).
  Two types of solutions have been analyzed. First, a class with constant redshift function and vanishing off diagonal elements in the Einstein tensor has been investigated.  The second category of solutions is related to non-constant redshift function.  Off diagonal elements will appear in this class of solutions, which can be compensated for, by extra terms resulting from the brane scenario. The brane-world scenario helps us to consider this class of solutions. It was shown that off diagonal terms in the right hand side of Einstein filed equations could be considered as the consequence of local bulk effects.

For constant redshift function, $\rho$ and $p$ are only functions of time, and are not space-dependent. So the Ricci scalar is only time dependent. It was shown that, spatial-Schwarzchild is the only asymptotically flat  possible solution for constant redshift function. The solutions for energy density, pressure and scale factor have been presented for different forms of Ricci scalar. It was shown that vacuum solution is an special case of vanishing Ricci scalar solution. Also, constant energy density is a special case of constant Ricci scalar solutions.
The NEC is satisfied for positive energy density. So, it is a good candidate avoiding exotic matter in wormhole theory. We have seen that for this class of solutions, time evolution of the scale factor is $a(t)\varpropto t^{1/2}$ which is similar to radiation FRW flat model.
Solutions with constant $R$ have exponential time dependence $a(t)$. The general behavior of energy density is approximately the same as vanishing $R$  solutions.  Energy density and pressure tend to constant values as $t\rightarrow \infty$  instead of zero. It was shown that vacuum and constant solutions are the special case of vanishing or constant Ricci scalar which can be archived by choosing the constant of integrations in Eqs. (\ref{20b}) and (\ref{20d}) equal to zero.

In the second category of solutions, nonconstant redshift function has been considered. Solutions with $p=-\rho$ is the only possible choice for $R=0$ in this category. Two exact solutions have been presented in detail for $R=0$ and nonconstant redshift function. The behavior of non-local bulk effects are different for these two solutions. The only possible solutions for constant Ricci scalar is related to constant redshift function, so there are no solutions with nonconstant redshift function and constant Ricci scalar. This is an important point in studying wormholes in the brane-world scenario.  It was shown that nonconstant redshift function solutions satisfy NEC. The validation region of NEC  for constant $U$ is dependent of the value of $R$ and the sign of energy density solutions. Although energy density has two solutions correspond to $\pm$ signs, it was shown that $\rho_+$ is a better candidate to describe wormhole in the brane-world scenario. Since solutions with $R=cte$ and $U=1$ satisfy NEC when $R\geq 0$, a positive Ricci scalar is positive Ricci scalar is the only acceptable choice.
 The EoS of matter for the nonconstant solutions is a linear form with $\omega=-1$. This equation is asymptotically linear for constant redshift function solutions. The asymptotically effective EoS parameter is  $\omega_{eff}(\infty)=-1$ or $\omega_{eff}(\infty)=1/3$. For time-dependent $R$, EoS parameter  is dependent on the constant of integration . Because solutions with  time-dependent $R$  and $U=1$ have a variable $\omega_{eff}(\infty)$ which is dependent to constant of integration, and have a vanishing $R(\infty)$, $\rho(\infty)$ and $p(\infty)$.  This kind of solutions seem to be better candidates in contrast to the other form of Ricci scalar to present wormhole solutions in the brane-world scenario.

In this article, the possibilities and restrictions on finding exact dynamic wormhole solutions within the brane-world scenario were studied. Some consequences and priorities of brane-world scenario in contrast to ordinary GR theory were discussed. Avoiding violation of the NEC, as one of the main ingredients of wormhole theory, is facilitated in brane-world dynamic wormholes. Since the physics of wormhole theory is based on theoretical studies, the study of wormhole theory in modified theories like brane-world may open a new window to the exploration of this amazing idea. Also, this study will help us to investigate  the power of brane-world scenario to explain the phenomena  which is  not completely understandable in ordinary GR.
.

\end{document}